%% file: preprint.tex
\documentclass{article}

\PassOptionsToPackage{numbers, compress}{natbib}

 \usepackage[preprint]{neurips_2026}

\usepackage[utf8]{inputenc} 
\usepackage[T1]{fontenc}    
\usepackage{url}            
\usepackage{booktabs}       
\usepackage{amsfonts}       
\usepackage{nicefrac}       
\usepackage{microtype}      
\usepackage{amsthm}
\usepackage{amsmath}
\usepackage{tabularx}
\usepackage{titletoc}

\usepackage[usenames, dvipsnames]{color} 
\usepackage{enumerate, amsmath, amsthm, amssymb, algorithm, algpseudocode, pifont, csquotes, dashrule, tikz, bbm, booktabs, bm, verbatim, url, dsfont}
\usepackage[framemethod=TikZ]{mdframed}
\usepackage{natbib}
\definecolor{princetonorange}{rgb}{0.90588235294, 0.45882352941, 0}
\definecolor{tiwagreen}{HTML}{007F66}
\usepackage[colorlinks]{hyperref}
\hypersetup{
     colorlinks = true,
     linkcolor = blue,
     anchorcolor = blue,
     citecolor = blue,
     filecolor = blue,
     urlcolor = blue
     }
\usepackage{caption}
\usepackage{subcaption}
\usepackage{authblk}
\usepackage{array}
\usepackage{xspace}
\usepackage{multirow}
\usepackage{enumitem}
\usepackage{wrapfig}

\theoremstyle{definition}
\newtheorem{definition}{Definition}

\newcolumntype{L}[1]{>{\raggedright\arraybackslash}m{#1}}

\title{Improving the Efficiency of Language Agent Teams \\ with Adaptive Task Graphs}

\author{%
  Elizabeth Mieczkowski$^{1}$\thanks{Code: \url{https://github.com/emieczkowski/latte}} \quad
  Alexander Ku$^{1}$ \quad
  Tiwalayo Eisape$^{1}$ \quad
  Dilip Arumugam$^{1}$ \\
  \vspace{-0.5em}
  John Matters$^{1}$ \quad
  Katherine M. Collins$^{1,2,3}$ \quad
  Ilia Sucholutsky$^{4}$ \quad
  Thomas L. Griffiths$^{1}$
   \vspace{0.5em}
  \\
  {\small
  $^{1}$Princeton University \quad
  $^{2}$University of Cambridge \quad
  $^{3}$MIT \quad
  $^{4}$New York University \\
  }
}

\input{commands}

\newif\ifsubmit
\submitfalse
\ifsubmit
\newcommand{\dnote}[1]{}
\newcommand{\enote}[1]{}
\newcommand{\tnote}[1]{}
\else
\newcommand{\dnote}[1]{\textcolor{blue}{Dilip: #1}}
\newcommand{\enote}[1]{\textcolor{orange}{Elizabeth: #1}}
\newcommand{\tnote}[1]{\textcolor{tiwagreen}{Tiwa: #1}}

\fi

\definecolor{princetonorange}{rgb}{0.90588235294, 0.45882352941, 0}

\begin{document}

\maketitle


\begin{abstract}
  Large language models (LLMs) are increasingly deployed in teams, yet existing coordination approaches often occupy two extremes. Highly structured methods rely on fixed roles, pipelines, or task decompositions assigned a priori. In contrast, fully unstructured teams enable adaptability and exploration but suffer from inefficiencies such as error propagation, inter-agent conflicts, and wasted resources (measured in time, tokens, or file operations). We introduce Language Agent Teams for Task Evolution (\textbf{LATTE}), a framework for coordinating LLM teams inspired by distributed systems, where processors must operate under partial observability and communication constraints. In LATTE, a team of agents collaboratively construct and maintain a shared, evolving coordination graph which encodes sub-task dependencies, individual agent assignment, and the current state of sub-task progress. This protocol maintains consistency while empowering agents to dynamically allocate work, adapt coordination, and discover new tasks. Across multiple collaborative tasks and a variety of base models, we demonstrate how LATTE reduces token usage, wall-clock time, communication, and coordination failures (e.g. file conflicts and redundant outputs) while matching or exceeding the accuracy of standard designs including MetaGPT, decentralized teams, top-down Leader-Worker hierarchies, and static decompositions.
\end{abstract}

\section{Introduction}

Collaboration can empower groups to achieve tremendous feats \citep{tomasello2005understanding, henrich2015secret}, but it comes with substantial coordination costs. In complex domains such as software development, distributing work across a team can outperform even the most skilled programmer \citep{conway1968committees, williams2000strengthening}, but also incurs substantial overhead \citep{brooks1975mythical}. What happens when one collaborator modifies a core function that others depend on? In what order should interdependent tasks be executed? How should work be allocated when team members differ in speed or reliability? And how can teams prevent local errors from cascading through the system? In practice, these challenges arise routinely, including concurrent edit conflicts \citep{ghiotto2020nature}, super-linear communication overhead \citep{brooks1975mythical}, dependency misalignment \citep{cataldo2008socio, maccormack2012exploring}, delays from distribution \citep{herbsleb2003empirical}, heterogeneous productivity \citep{sackman1968exploratory}, and stragglers \citep{dean2013tail}. Collaboration can dramatically amplify capability, but only when coordination is effectively managed.

Recent work has shown that LLM teams can improve accuracy and problem-solving by distributing tasks, roles, and context across multiple agents
\cite{anthropic2025multiagent, bhattacharyyasocial, li2024more, swanson2025virtual, wu2024autogen, zhang2024chain}, demonstrating emergent coordination~\cite{park2023generative}.
Despite these successes, the design of LLM teams remains fundamentally limited. Modern LLMs derive much of their power from their ability to flexibly adapt to new contexts and tasks \cite{dong2024survey}. In contrast, most existing LLM team architectures impose rigid coordination structures, assigning fixed roles or pre-specifying task decompositions prior to execution \cite{hong2023metagpt, qian2024chatdev}. We argue that constraining LLM teams in this way limits their capacity for dynamic adaptation \cite{riedl2025emergent} and introduces fragility, whereby failures or hallucinations in the Lead propagate downstream \cite{jo2025byzantine}.

Unstructured or decentralized LLM teams are more flexible \cite{liu2026learning}, yet face their own challenges. Without coordination scaffolding, agents frequently overwrite one another, produce inconsistent or incorrect outputs, and erroneously report task completion \cite{mieczkowski2026language, shapira2026agents}. These failures worsen when tasks must be performed sequentially, where adding agents to a team leads to over-communication and performance degradation \cite{kim2025towards, mieczkowski2026language}. Self-coordinating teams often cannot outperform single expert models \cite{pappu2026multi} and are unable to overcome failures propagated by individual agents \cite{berdoz2026can}.
This suggests a \textit{fundamental tension in designing LLM teams}: structure is needed to improve coordination and inter-agent consistency, but over-specification suppresses the adaptability that makes LLMs powerful. 

\begin{figure}[t]
  \centering
  \includegraphics[width=\textwidth]{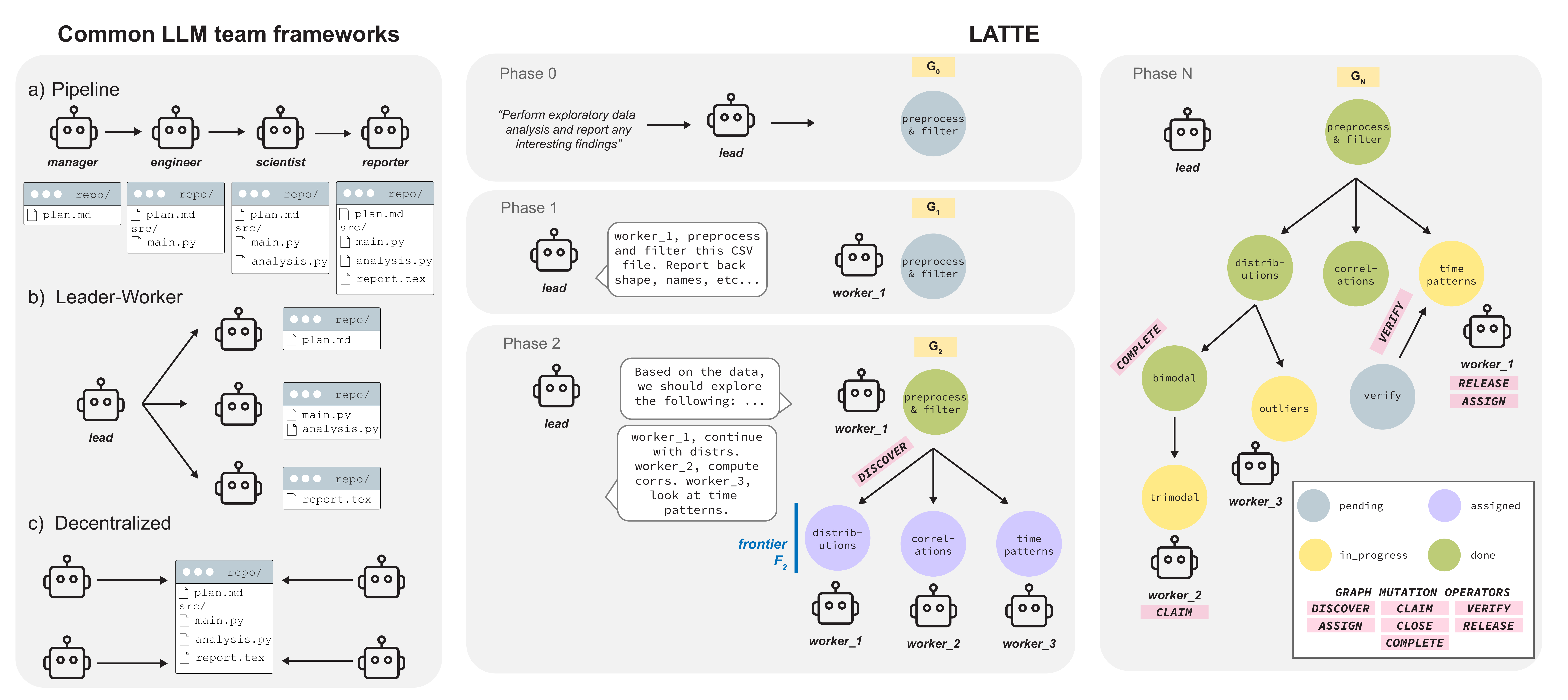}
  \caption{\textbf{LATTE.} Most existing LLM team designs are either highly structured (a. pipeline systems; b. Leader-Worker hierarchies) or unstructured (c. decentralized teams). (d) LATTE provides teams with a dynamic coordination graph that they collectively maintain and adapt. For example, in a data analysis task, the Lead initializes $G_0$ and assigns Worker 1 to preprocess. As Worker 1 learns about the data, it spawns parallel subtasks on the frontier (\textsc{Discover}) which the Lead then \textsc{ASSIGN}s to Workers 2 and 3. As this process continues, the Lead can \textsc{Release} stragglers and \textsc{Close} completed subtasks while Workers \textsc{Claim} frontier tasks proactively to avoid idleness. The shared graph serializes coordination decisions while preserving parallelism.}
  \label{fig:latte_fig1}
\end{figure}

To resolve this tension, we propose \textbf{LATTE} (\textbf{L}anguage \textbf{A}gent \textbf{T}eams for \textbf{T}ask \textbf{E}volution), a formal orchestration framework for LLM teams to explicitly represent and adapt their own coordination during execution (Figure~\ref{fig:latte_fig1}). Drawing inspiration from distributed systems --- where protocols enable reliable task scheduling under partial information and dynamic conditions \cite{agrawal2010executing, mao2019learning, woo1997task} --- agents operating within the LATTE framework construct and maintain a shared, evolving \textbf{coordination task graph}. In this graph, nodes represent subtasks that agents are assigned to, edges encode completion dependencies between subtasks, and a set of graph mutation operators allow the team to restructure coordination as execution unfolds. This graph serves as an evolving record of task decomposition, progress, roles, and active effort. Each agent proposes updates to the graph based on local context, which are reviewed and merged by a single agent. This division of labor reduces bottlenecks and single-points-of-failure that arise with hierarchical designs, provides explicit mechanisms for monitoring stragglers, and naturally serializes updates to prevent divergent local views or inconsistencies. The resulting framework preserves coherence and efficiency without sacrificing the capacity for emergent, context-sensitive adaptation.
Our contributions are as follows:
\begin{enumerate}[leftmargin=0.18in,topsep=0in]
    \item \textbf{A formal orchestration framework for LLM teams via dynamic task graphs:} We define a set of graph mutation operators (Discover, Assign, Claim, Complete, Release, Close, and Verify) with explicit preconditions, postconditions, and invariant-preservation guarantees which culminate in a rigorous execution protocol for multi-agent LLM coordination. We demonstrate that the graph structure induces desirable runtime properties such as maximal parallelism via frontier nodes. 
    \item \textbf{A hybrid centralized-decentralized model:} We introduce a two-tier coordination model in which worker agents propose structural modifications to the graph and a lead orchestrator accepts or rejects them, preserving global consistency while enabling local adaptability. This division of labor is grounded in a probabilistic account of task decomposition.
    \item \textbf{LLM team interpretability:} LATTE  externalizes coordination during task execution, providing ways to interpret and audit team behavior. Our evaluation provides a suite of coordination metrics (overwrite rate, concurrent conflicts, wasted characters, idle rounds, and straggler tail latency) that address a systematic gap in how multi-agent LLM systems are benchmarked.
    \item \textbf{Empirical validation:} LATTE consistently reduces token consumption, wall-clock time, inter-agent messages, file overwrites, consistency conflicts, and total output all while achieving higher accuracy than alternatively-structured and widely-used LLM team implementations. 
\end{enumerate}

Together, these results suggest that explicit coordination structures maintained by agents themselves are a viable path towards LLM teams that are simultaneously more efficient, interpretable, and adaptive.

\section{Related Work}

Prior work on LLM team coordination clusters around three patterns.
\textbf{Static} systems like MetaGPT and ChatDev assign fixed functional roles and task structures before execution begins \cite{hong2023metagpt, qian2024chatdev}. While this simplifies scheduling, static assignments may struggle when new dependencies emerge or workloads shift mid-execution. 
\textbf{Hierarchical or centralized} frameworks such as HuggingGPT and recent meta-agent approaches use a lead agent to plan, dispatch, and synthesize work across subordinates \cite{shen2023hugginggpt, liagent}. 
Centralization can enforce consistency but creates bottlenecks and single points of failure, which are especially acute in LLM 
teams where the Lead may hallucinate, crash, or fail to consolidate distributed progress \cite{shapira2026agents}.
\textbf{Decentralized} teams avoid bottlenecks by letting agents operate autonomously, improve diversity, and distribute long contexts across agents \cite{du2024improving, li2024more}. However, agents operating on local views of task state can produce conflicting or redundant outputs, and scaling the number of agents can degrade performance, particularly in tasks requiring sequential reasoning, expert agent assignment, or consistency \cite{kim2025towards, pappu2026multi, mieczkowski2026language}.

\textbf{Task graphs} from distributed computing offer a natural improvement to task decomposition and assignment: nodes represent tasks, edges encode precedence constraints, and schedulers assign work across processors efficiently \cite{topcuoglu2002performance, 
moritz2017ray}. Classic schedulers like HEFT compute globally optimized assignments before execution; dynamic variants and work-stealing approaches such as NABBIT assign tasks online as they become available \cite{johnson1993concurrent, agrawal2010executing}. However, these systems 
assume well-defined tasks and explicit control mechanisms. Extending task graphs to LLM teams requires supporting agents that autonomously discover, modify, and claim tasks in natural language. To our knowledge, LATTE is the first framework to bridge this gap: LLM teams jointly construct, maintain, and revise a shared task graph as an online, dynamic coordination structure during execution. We provide an extended discussion of related work in Appendix~\ref{app:related}.

\section{LATTE: Language Agent Teams for Task Evolution}

We establish four key desiderata for a structured LLM team execution framework motivated by the limitations of prior architectures.

\textbf{D1. Hybrid coordination:} To avoid the bottlenecks of fully centralized systems and the inconsistency of fully decentralized ones, coordination should be hybrid. Decisions affecting shared state (e.g., graph updates or artifacts) must be centrally mediated, while task execution should be opportunistic to allow for parallel progress.

\textbf{D2. Adaptive scaling:} The framework should deploy agents efficiently, dynamically activating agents based on the current workload while maximizing parallelism when dependencies allow.

\textbf{D3. Fault tolerance and monitoring:} Because agents may stall, hallucinate, or produce null outputs, the system must support active monitoring and dynamic reallocation to detect and reassign tasks from unresponsive agents. It should also support auditing, where agents can proactively identify and flag high-uncertainty outputs that warrant additional quality control.

\textbf{D4. Context scoping:} To prevent memory overload and confusion, each agent should receive a scoped context. Workers should see only their local subtask, while the Lead’s view is restricted to the coordination graph rather than the full execution history.

\subsection{Dynamic coordination graph}

The set of agents $\mathcal{A} = \{\ell\} \cup \mathcal{W}$ in a LATTE team belong to one of two types: a Lead $\ell$ responsible for maintaining coordination structure and a set of Workers $\mathcal{W}$ responsible for executing assigned subtasks. Coordination proceeds through a shared \textbf{dynamic coordination graph} that explicitly tracks task progress, agent assignments, and shared state as execution unfolds.

\begin{definition}[Dynamic Coordination Graph]
A \emph{dynamic coordination graph} $G_t$ at round $t \in \{1,\ldots,T\}$ is a directed acyclic graph $G_t = (V_t,\, E_t,\, \lambda_t)$. Here, $V_t$ is a finite set of nodes, each corresponding to a subtask. $E_t \subseteq V_t \times V_t$ is the set of dependency edges between subtasks such that $(u,v) \in E_t$ implies that subtask $v$ cannot begin until $u$ is complete. $\lambda_t : V_t \to (\mathcal{A} \cup \{\bot\}) \times S$ assigns each node an agent and a status, where $\bot$ denotes unassigned and $S := \{\texttt{pending},\ \texttt{assigned},\ \texttt{in\_progress},\
          \texttt{done}, \
          \texttt{verified}\}$.
\end{definition}

A strength of the coordination graph is encoding opportunities for parallelism during task execution. 

\begin{definition}[Frontier]
The \emph{frontier} $F_t \subseteq V_t$ at round $t$ is the set of pending nodes with no unsatisfied dependencies: $F_t := \{ v \in V_t \mid \text{status}(v) = \texttt{pending} 
    \text{ and } \forall (u,v) \in E_t, \text{status}(u) = \texttt{done} \}.$
\end{definition}

$F_t$ determines which subtasks are immediately executable, and thus corresponds to the number of Workers that can proceed in parallel at $t$.

\subsection{Graph mutation operators}

The asymmetry in information between Lead and Workers directly determines operator permissions. Worker $w_i$, reasoning from local trace $d_t^{(i)}$ about its subtask $v$, has sufficient information to propose local changes, such as discovering new subtasks encountered during execution (\textsc{Discover}) and certifying its subtask's completion (\textsc{Complete}). It lacks visibility beyond $v$, preventing safe and holistic evaluation of proposals with graph-wide consequences, such as forcing completion of nodes whose downstream effects it cannot observe (\textsc{Close}). The Lead $\ell$ maintains global visibility into $G_t$ and exclusively controls operators with graph-wide consequences, such as \textsc{Release} to reassign stalled work and \textsc{Verify} to intercept errors. 

Unlike the other operators, task acquisition need not be centralized. Rather than requiring $\ell$ to \textsc{Assign} every subtask, idle Workers may proactively claim available work directly from $v \in F_t$ via \textsc{Claim}($v$). This mirrors work-stealing or self-scheduling principles in distributed computing \cite{polychronopoulos2009guided, agrawal2010executing}, where fast processors claim ready tasks from a shared queue rather than waiting on a central scheduler, reducing overhead and improving wall-clock time. Concurrent claims on the same $v \in F_t$ are resolved by the orchestrator, preserving serialization without centralizing acquisition.

\begin{wraptable}{r}{0.7\textwidth}
\centering
\vspace{-0.35cm}
\resizebox{\linewidth}{!}{
\begin{tabular}{l c l}
\toprule
\textbf{Operator} & \textbf{Caller} & \textbf{Output} \\
\midrule
$\textsc{Discover}(v, \text{deps})$ & $\ell, w$ & Add \texttt{pending} node $v$ with \texttt{deps} \\
$\textsc{Assign}(v, w)$            & $\ell$    & Assign pending $v$ to Worker $w$ \\
$\textsc{Claim}(v)$            & $w$    & Worker $w$ claims node $v$ \\
$\textsc{Complete}(v)$             & $w$       & Mark $v$ as finished by its $w$ \\
$\textsc{Release}(v)$              & $\ell$    & Return $v$ to \texttt{pending} \\
$\textsc{Close}(v)$                & $\ell$    & Force-complete $v$ \\
$\textsc{Verify}(v)$               & $\ell$    & Spawn verification for $v$ \\
\bottomrule
\end{tabular}
}
\caption{Graph mutation operators.}
\label{tab:operators}
\vspace{-0.2cm}
\end{wraptable}

Accordingly, LATTE provides a set of graph mutation operators with explicit preconditions, postconditions, and invariant-preservation guarantees (Table~\ref{tab:operators}, Appendix~\ref{app:operators_app}). Unlike 
prior multi-agent LLM frameworks, where coordination contracts are implicit in role prompts, LATTE makes these contracts explicit and verifiable for every operator.

\textbf{DAG invariance.} All operators preserve the DAG invariant on $G_t$. \textsc{Discover} is the 
only operator that adds edges to $G_t$, and its precondition $v \notin V_t$ guarantees no self-loops. Additionally, requiring \texttt{deps} $\subseteq V_t$ with the resulting graph acyclic ensures no cycles are introduced. All remaining operators modify only $\lambda_t$ and leave $(V_t, E_t)$ unchanged.

\textbf{Probabilistic motivation.} This Leader-Worker division of labor can also be interpreted through a probabilistic lens, viewing the dynamic evolution of the task graph as an approximate posterior inference problem. To enable context scoping, LATTE decouples the graph updating process into proposal (by the Workers) and evaluation (by the Lead), which is conceptually grounded in sampling procedures such as Metropolis-Hastings. Further details of this motivation can be found in Appendix~\ref{app:prob_motiv}.

\subsection{LATTE execution}

The full execution protocol is described in Algorithm~\ref{app:latte_exec}. LATTE proceeds in two phases: a preliminary planning phase followed by an iterative execution loop over the dynamically evolving coordination graph. During planning, $\ell$ is given the task description and initializes $G_0$ by proposing an initial decomposition of the problem via \textsc{Discover} operations. During execution, at each round $t=1,...,T$, agents are selectively dispatched to operate on the current graph $G_t$. Each round consists of five steps: (1) heartbeat monitoring (i.e., periodic liveness checks) to flag stragglers or stalled Workers to $\ell$; (2) frontier identification to compute $F_t$, the queue of available tasks; (3) dispatching for agents that are in-progress, assigned to new frontier tasks, and $\ell$ when necessary; (4) parallel execution of all selected agents; and (5) termination, which returns all task outputs and $G_t$. 

\subsection{Coordination properties}

The LATTE protocol equips LLM teams with the structural tools to satisfy the four design desiderata (D1-D4) introduced in Section 3.1.

\textbf{D1. Hybrid coordination:} To balance consistency with adaptability, the Lead $\ell$ maintains exclusive control over operators with graph-wide consequences (\textsc{Assign}, \textsc{Verify}, \textsc{Close}), while Workers $w$ operate autonomously within their local scope. Crucially, LATTE enables \textit{self-scheduling}: idle Workers may invoke \textsc{Claim}($v$) for any $v \in F_t$, allowing for opportunistic execution that reduces Lead overhead and improves wall-clock time \cite{polychronopoulos2009guided}. All structural updates to $G_t$ are serialized, preventing race conditions and inconsistencies to shared state. 

\textbf{D2. Adaptive scaling:} At each round $t$, the number of Workers dispatched by LATTE equals $\min(|F_t|, |\mathcal{W}|)$. This ensures maximal parallelism given $G_t$; no valid protocol can dispatch more Workers at $t$ without violating a dependency constraint. Agents are activated only when there is available work, eliminating idle computation.

\textbf{D3. Fault tolerance and monitoring:} The heartbeat mechanism flags Workers that have been assigned but remain inactive for $H$ rounds, surfacing potential stalls to $\ell$ before they block progress. Upon detecting a straggler, $\ell$ may invoke \textsc{Release}($v$) to return the task to a pending state, making it available for immediate re-assignment or self-scheduling. In addition, rather than mandating an expensive review of every subtask, LATTE supports emergent verification. The \textsc{Verify} operator is invoked selectively by $\ell$ on nodes judged to be high-risk or high-uncertainty. This allows the rigor of quality control to scale dynamically with the complexity of the task graph.

\textbf{D4. Context scoping:} To mitigate context accumulation and token exhaustion, Workers receive only the description of their assigned subtask and its direct predecessors. The Lead receives $G_t$ and agent messages but does not ingest individual Workers' full execution traces. These constraints bound the context each agent must attend to, reducing reasoning errors caused by irrelevant information.

\section{Experiments}
We evaluate LATTE against several existing multi-agent frameworks across three newly-designed collaborative domains: exploratory data analysis, debugging, and code generation. These domains were chosen to stress-test different coordination properties (parallelism, consistency, and adaptability). Experimental settings and prompt designs are provided in Appendices~\ref{app:latte_impl} and \ref{app:baselines}.
Full task specifications and evaluation criteria are provided in Appendix~\ref{app:experiments}.

Each task rewarded different combinations of the coordination properties above:

\textbf{Task 1: Exploratory Data Analysis.} Agents performed exploratory analysis on an opaque dataset, requiring preprocessing, analysis, visualizations, and synthesis of findings. Correctness was evaluated by a private test suite checking whether agents correctly identified planted data properties.

\textbf{Task 2: Debugging.} Agents debugged an existing repository against a test suite, requiring iterative test execution and code modification. This task rewards both parallelism (independent bugs can be diagnosed 
simultaneously) and consistency (some functions can only be verified after dependencies are fixed). We placed several bugs in a signal-processing library, and success required teams to pass all tests in a given suite.

\textbf{Task 3: Library Extension.} Agents extended a Python text-processing library by completing two existing classes and building six new modules from stubs. The task has natural sequential dependencies, parallel modules, and a final integration step. Correctness was evaluated by a private test suite after completion. Unlike Tasks 1 and 2, the required functions are fully known in advance. 

\begin{figure}[t]
  \centering
  \includegraphics[width=0.85\textwidth]{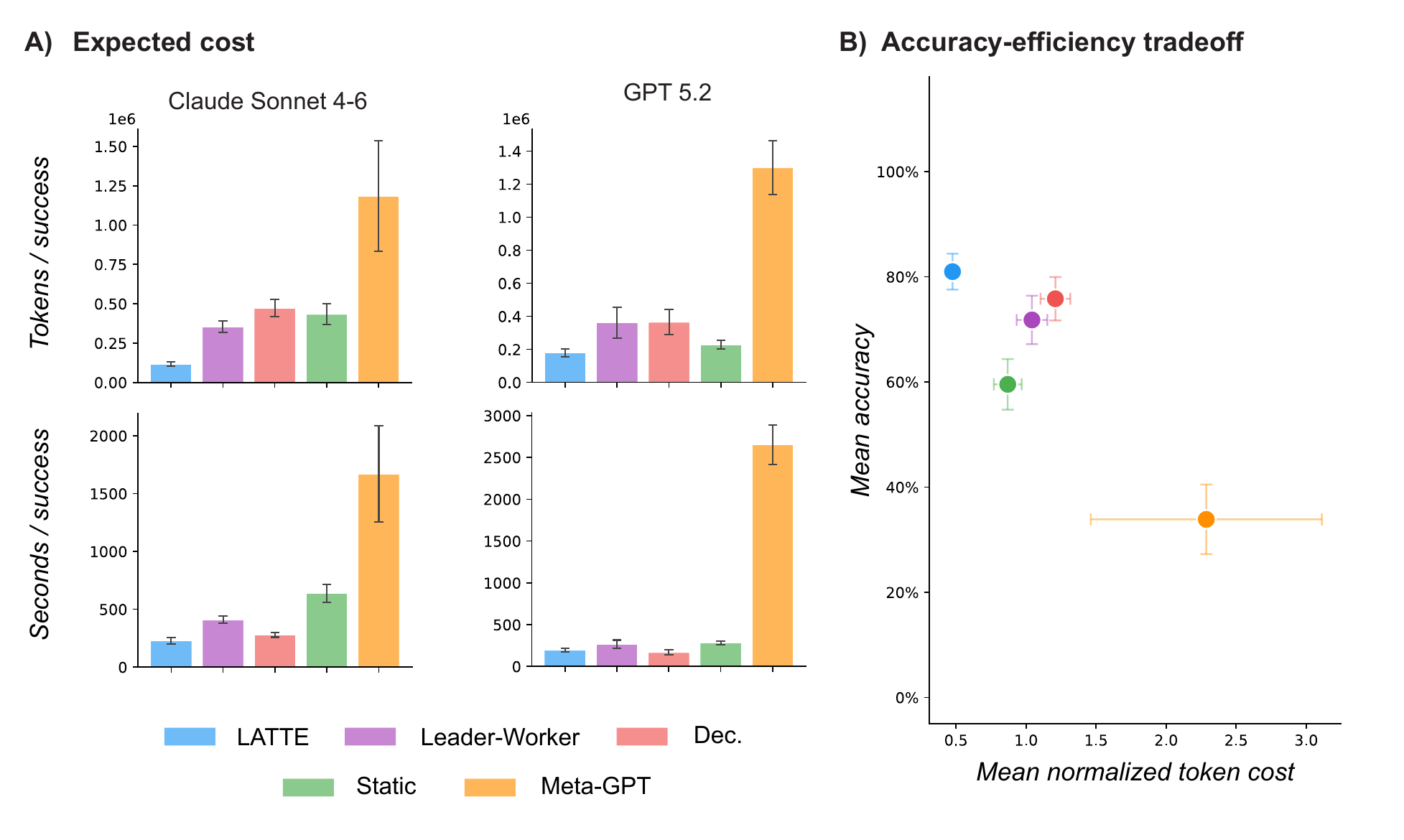}
  \caption{\textbf{Efficiency-accuracy tradeoff.} A) LATTE achieves greater efficiency than alternative frameworks. We measure \textit{expected} cost (total tokens or wall-clock time weighted by trial completion rate) to account for runs in which teams fail to terminate. B) LATTE achieves higher task success with lower token consumption (normalized across tasks) on the accuracy-vs-token-cost Pareto frontier.
  }
  \label{fig:pareto}
\end{figure}

We evaluate the performance of LATTE against four baseline team structures. We test Leader-Worker hierarchies, where a single Lead synthesizes and assigns tasks to four Workers; MetaGPT \cite{hong2023metagpt}, representative of pipeline-based LLM teams with different roles (Product Manager, Architect, Project Manager, Engineer, QA Engineer); decentralized teams with 5 peer agents; and a static task graph ablation, in which the Lead initializes assignments based on its prior $G_0$ and agents cannot update the graph after planning. We maintained a team size of $N=5$ to benchmark against MetaGPT. For each team structure, we tested two frontier base models: Claude Sonnet 4-6 (Anthropic; \texttt{claude-sonnet-4-6}) and GPT-5.2 (OpenAI; \texttt{gpt-5.2}). We ran 10 trials per condition for a total of 300 trials (5 conditions × 2 models × 3 tasks × 10 repetitions). Full implementation details are provided in Appendix~\ref{app:latte_impl} and \ref{app:baselines}.

\section{Results}

\subsection{LATTE achieves higher accuracy and efficiency than existing LLM teams}

\begin{table}[t]
\centering
\begin{small}
\setlength{\tabcolsep}{3pt}
\renewcommand{\arraystretch}{0.8}
\caption{Accuracy, token usage, and wall-clock time across successful trials 
per task ($\pm$ SEM).}
\label{tab:results}
\begin{tabular}{ll ccccc}
\toprule
& & \textbf{LATTE} & \textbf{Leader-Worker} & \textbf{Decentralized} 
& \textbf{Static} & \textbf{MetaGPT} \\
\midrule
\multirow{4}{*}{\textit{Acc. (\%)}}
& Agg.          & $80 \pm 4$   & $70 \pm 5$   & $74 \pm 4$   & $58 \pm 5$   & $34 \pm 7$ \\
& Data Analysis & $96 \pm 1$   & $94 \pm 1$   & $93 \pm 2$   & $88 \pm 3$   & $75 \pm 2$ \\
& Debug         & $100 \pm 0$  & $90 \pm 7$   & $100 \pm 0$  & $44 \pm 13$  & $32 \pm 11$ \\
& Library Ext.  & $40 \pm 2$   & $23 \pm 2$   & $27 \pm 2$   & $40 \pm 2$   & $6 \pm 4$ \\
\midrule
\multirow{4}{*}{\textit{Tokens (K)}}
& Agg.          & $148 \pm 14$ & $379 \pm 51$ & $419 \pm 47$ & $297 \pm 40$ & $397 \pm 59$ \\
& Data Analysis & $122 \pm 13$ & $257 \pm 60$ & $271 \pm 60$ & $403 \pm 93$ & $390 \pm 61$ \\
& Debug         & $227 \pm 33$ & $642 \pm 103$& $792 \pm 73$ & $286 \pm 36$ & $236 \pm 41$ \\
& Library Ext.  & $98 \pm 9$   & $169 \pm 17$ & $194 \pm 25$ & $140 \pm 12$ & $707 \pm 188$ \\
\midrule
\multirow{4}{*}{\textit{Wall-clock (m)}}
& Agg.          & $3.5 \pm 0.3$  & $5.9 \pm 0.6$  & $3.7 \pm 0.3$  & $6.0 \pm 0.6$  & $11.5 \pm 1.2$ \\
& Data Analysis & $3.2 \pm 0.3$  & $4.9 \pm 0.9$  & $2.9 \pm 0.5$  & $6.2 \pm 0.6$  & $8.7 \pm 1.7$ \\
& Debug         & $5.3 \pm 0.6$  & $9.1 \pm 1.2$  & $6.1 \pm 0.4$  & $6.2 \pm 0.6$  & $8.7 \pm 1.7$ \\
& Library Ext.  & $2.1 \pm 0.2$  & $3.6 \pm 0.3$  & $2.2 \pm 0.2$  & $3.4 \pm 0.3$  & $18.9 \pm 3.4$ \\
\bottomrule
\end{tabular}
\end{small}
\end{table}

Across tasks and base models, LATTE consistently achieves a superior accuracy–efficiency tradeoff, Pareto-dominating existing LLM team structures (Fig.~\ref{fig:pareto}; Table~\ref{tab:results}). 

\textbf{Computational cost.} Using one-sided Mann-Whitney U tests on normalized, pooled costs across all tasks and models, LATTE achieves a mean token cost of $47.5\%$, nearly half that of the next-best method, the static graph ablation ($M=86.9\%$, $p<0.01$). All other baselines are also more expensive: MetaGPT ($M=228.7\%$, $p<0.01$), Leader-Worker ($M=104.2\%$, $p<0.01$), and decentralized ($M=120.9\%$, $p<0.01$).

\textbf{Wall-clock time.} LATTE ($M=66.7\%$) is faster than static graphs ($M=110.7\%$, $p<0.01$), MetaGPT ($M=289.0\%$, $p<0.01$), and Leader--Worker teams ($M=105.7\%\%$, $p<0.01$). Decentralized teams also have a higher mean latency than LATTE, though the difference is not statistically significant ($M=69.3\%$, $p=0.34$). 

\textbf{Task accuracy.} LATTE achieves the overall highest task accuracy ($79.7\%$), surpassing static graphs ($57.6\%$, $p<0.01$), fixed pipelines (MetaGPT; $33.9\%$, $p<0.01$), Leader-Worker teams ($70.1\%$, $p=0.04$), and decentralized teams ($73.9\%$, $p=0.16$). Consistent with our probabilistic motivation, LATTE achieves comparable accuracy to the static ablation on Task 3 (using substantially less wall-clock time and fewer tokens) but much greater accuracy on Tasks 1–2. When task structure is known in advance, the initial graph $G_0$ can be well-specified, leaving fewer subtasks for LATTE to dynamically discover. 

\subsection{LLM teams successfully utilize dynamic coordination graphs via LATTE}

\begin{figure}[t]
  \centering
  \includegraphics[width=.8\textwidth]{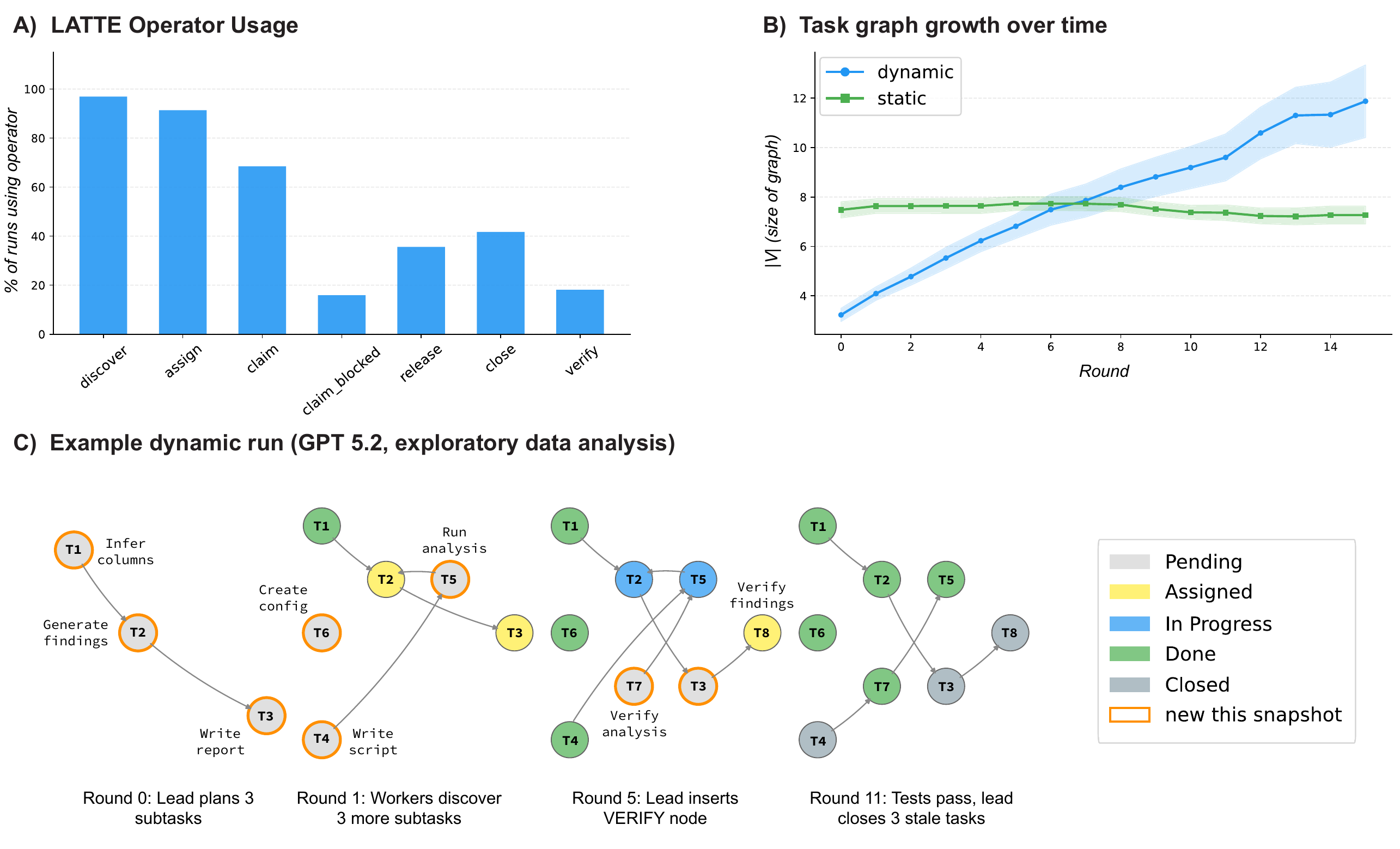}
  \caption{\textbf{LLM teams successfully utilize LATTE.} A) LATTE teams emergently call all graph operators across rounds, demonstrating full utilization of the coordination toolkit. B) Dynamic coordination graphs grow larger than static ones over time. This reflects richer and more fine-grained understanding of which subtasks need to be executed, offering more opportunities for Workers to be deployed. In contrast, a smaller static graph represents a fixed, underspecified coordination structure that cannot adapt to task demands. C) A representative example of how a coordination graph evolves across rounds using GPT-5.2. As agents add edges and dependencies, the graph encodes emergent assignments and progress. Because coordination is represented explicitly, team behaviors can be tracked, interpreted, and diagnosed over time, a key advantage over black-box multi-agent systems. }
  \label{fig:latte_operator}
\end{figure}

Figure~\ref{fig:latte_operator} demonstrates that LLM teams successfully utilize the full expressive power of the LATTE protocol to manage task complexity. \textsc{Discover} is the most frequent operator, confirming that teams actively expand their coordination graphs as new requirements emerge. The Lead effectively delegates via \textsc{Assign}, yet the high success rate of Worker-initiated \textsc{Claim} operations suggests that decentralized self-scheduling significantly reduces coordination bottlenecks. 
Notably, LATTE teams exhibit emergent fault tolerance through the selective use of recovery operators. The Lead invoked \textsc{Release} in 36\% of trials to reassign straggling tasks. Similarly, \textsc{Verify} was invoked in 19\% of trials, demonstrating that the Lead can trigger verification when deemed necessary. 
Specifically, Leads triggered \textsc{Verify} more often in high-uncertainty, challenging trials, where teams took an average of 18.1 rounds to pass tests. In contrast, trials that completed successfully in 8.1 rounds on average contained no verification events.
These behaviors are particularly encouraging as they indicate LATTE's capacity for autonomous fault tolerance, monitoring, and strategic resource allocation. 

Crucially, these patterns are only observable because LATTE \textit{externalizes} emergent coordination, providing concrete ways to interpret and audit team behavior. The evolving task graph makes otherwise hidden decisions directly observable. For example, Figure~\ref{fig:latte_operator}C demonstrates how team structure and progress evolve over time. In a Leader-Worker or decentralized team, mechanisms like selective verification would be invisible or buried in the message logs rather than recorded as explicit and structured coordination decisions. 

\subsection{LATTE induces better coordination}

Beyond aggregate performance, we report finer-grained coordination metrics to address a fundamental gap in evaluating teams: overwrites, concurrent writes, communication overhead, and wasted outputs are rarely measured in prior work, yet directly capture how often agents waste resources and interfere. 

\begin{figure}[t]
  \centering
\includegraphics[width=0.8\textwidth]{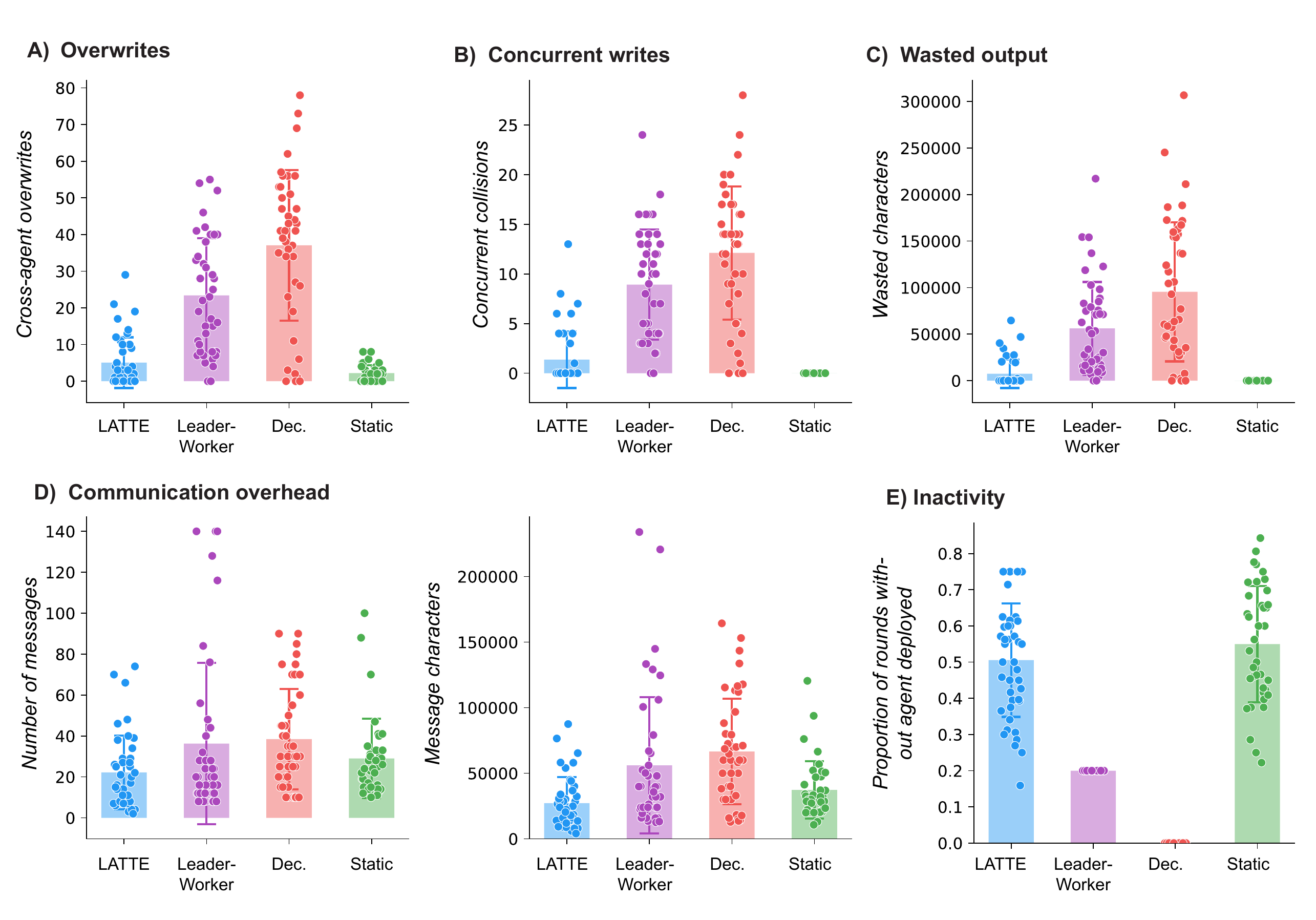}
  \caption{\textbf{LATTE improves coordination.} (A)~\textit{Overwrites}: agents overwriting a prior agent's work in a later round. (B)~\textit{Concurrent writes}: two agents simultaneously writing to the same function. (C)~\textit{Wasted output}: characters written that do not appear in the final output. (D)~\textit{Communication overhead}: number and volume of messages exchanged. (E)~\textit{Inactivity}: proportion of rounds with an agent suppressed. LATTE reduces A--D and increases E.
  }
  \label{fig:consistency}
\end{figure}

\textbf{LATTE decreases inter-agent conflicts.}  In collaborative tasks with a shared state, agents may overwrite completed work, simultaneously edit the same text, or redundantly duplicate effort. Each conflict carries distinct costs in wasted tokens, corrupted state, and downstream debugging. LATTE mitigates these failure modes through explicit task assignment and dependency tracking, ensuring agents operate on disjoint subtasks in a well-defined order (Fig.~\ref{fig:consistency}A-C).
The reductions are substantial. LATTE agents overwrite each other $4.3\times$ per trial on average, versus $22.8\times$ in Leader-Worker teams ($p<0.01$) and $35.4\times$ in decentralized ($p<0.01$): a $5.3\times$ and $8.2\times$ reduction respectively. Concurrent writes to the same function follow the same pattern: $1.0\times$ per trial versus $8.5\times$ in Leader-Worker ($p<0.01$) and $11.5\times$ in decentralized ($p<0.01$). These conflicts compound into wasted output. LATTE produces 5,236 discarded characters per trial on average, compared to 45,436 in Leader–Worker ($p<0.01$) and 78,062 in decentralized ($p<0.01$), corresponding to more than 40,000 and 70,000 extra characters of output that never appear in the final product. 

\textbf{LATTE reduces costly communication.}
Communication overhead is also a meaningful cost in LLM teams. Agents tend to send excessive messages, consuming unnecessary tokens, introducing latency, and interrupting teammates with irrelevant information \cite{mieczkowski2026language, zhang2024cut, cemri2025multi}. LATTE constrains this by making communication purposeful (Fig.~\ref{fig:consistency}D). 
Workers message the Lead only when blocked, needing clarification, or signaling dependencies for another agent.
This structure produces measurably less communication. LATTE agents send $20.4$ inter-agent messages per task, compared to $31.4$ in Leader–Worker teams 
($p=0.04$) and $34.8$ in decentralized teams ($p<0.01$). LATTE agents also exchange leaner messages, with $42{,}484$ characters sent versus $50{,}073$ in Leader-Worker teams ($p<0.01$) and $60{,}394$ in decentralized teams ($p<0.01$). These results show that LATTE reduces unnecessary communication and limits context accumulation per agent.

\textbf{LATTE selectively activates agents, reducing idle computation.}
LATTE restricts participation to $F_t$, activating agents only when pending work exists. On average, agents are active for only $48.7\%$ of rounds while maintaining high task performance (Fig.~\ref{fig:consistency}E). By contrast, decentralized teams activate all agents every round regardless of demand, inflating computation and communication costs. Leader-Worker teams fall in between, deploying agents for $80\%$ of rounds.

\textbf{LATTE mitigates stragglers.}
Finally, a key challenge in static systems are stragglers: agents that take disproportionately long to complete assigned tasks \cite{dean2008mapreduce, mieczkowski2026language}.
LATTE addresses this by monitoring node execution time and sending a heartbeat to the Lead when a threshold is exceeded, giving the Lead the option to \textsc{Release} and reassign the task, or Workers the option to self-claim it. \textsc{Release} is invoked in $36\%$ of runs, confirming that straggler mitigation emerges in practice. The impact on completion time is substantial. LATTE teams complete assigned nodes in 39.2s on average versus 75.6s for static teams ($p<0.01$), and this gap widens at the tail: at the 95th percentile, 130s versus 294s for static teams (a $2.3\times$ reduction), showing this mechanism effectively prevents stragglers from blocking task completion.

\section{Conclusion}

Inspired by task graphs and scheduling protocols in distributed systems, LATTE enables LLM teams to dynamically explore and solve problems while maintaining consistency and efficiency. Across settings, LATTE exhibits consistent gains in token and time efficiency, conflict resolution, and coordination quality while preserving the flexibility for agents to explore and refine solutions as tasks evolve. More broadly, LATTE challenges a core assumption in prior frameworks that coordination structure must be imposed by the system architect. Instead, allowing agents to maintain and revise their own coordination structure online empowers teams to adapt to evidence accumulated during execution. These results establish explicit, agent-maintained coordination as a principle for building LLM teams that are simultaneously more efficient, reliable, and adaptive. By reducing token consumption and wall-clock time, LATTE directly cuts the computational (and thereby financial) cost associated with LLM teams, conserving resources that would otherwise be spent on redundant communication and unresolved conflicts.

\textbf{Limitations.} Several limitations point to promising directions for future work. First, LATTE introduces planning overhead from graph initialization which may outweigh its benefits on short or simple tasks where a single agent would suffice. A deeper analysis of this overhead is provided in Appendix~\ref{app:graph_overhead}. Second, our evaluation focuses on domains with natural subtask boundaries such as coding. The current operator set could be extended to support less structured forms of task discovery and decomposition, enabling LATTE to tackle a broader range of open-ended reasoning tasks in future work. Third, we fix team size to benchmark fairly against baselines such as MetaGPT, leaving the question of how LATTE scales to future work. Fourth, LATTE teams demonstrate an emergent ability to identify where verification is needed, invoking \textsc{Verify} selectively rather than applying blanket review mechanisms to all agent outputs \cite{huang2024resilience}. Understanding and strengthening these emergent verification mechanisms is a particularly promising direction, as targeted quality control may be key to reliable LLM team performance at scale. Finally, future work should explore fine-tuning or reinforcement learning on coordination outcomes, allowing teams to learn better graph constructions, task assignments, and communication protocols over time. 

\textbf{Code Availability.}
An implementation of LATTE is available at \url{https://github.com/emieczkowski/latte}. 

\section*{Acknowledgments} 

This work was supported by the National Defense Science and Engineering Graduate (NDSEG) Fellowship Program to EM, and ONR MURI N00014-24-1-2748 to DA and TG. 

\bibliographystyle{plainnat}
\bibliography{references}


\newpage
\setcounter{section}{0}
\appendix  
\renewcommand{\thesection}{A\arabic{section}}

\startcontents[appendix]
\printcontents[appendix]{l}{1}{\section*{Appendix}}

\section{Related Work}
\label{app:related}

\subsection{LLM teams}

LLMs are increasingly deployed in multi-agent teams. In some settings, these teams outperform individual models by improving diversity and distributing long contexts across many agents \cite{zhang2024chain, li2024more, bhattacharyyasocial, du2024improving}. As a result of these emergent cooperative abilities, LLM teams have achieved desirable results in domains such as scientific discovery \cite{swanson2025virtual, zhangposition} and software engineering \cite{hong2023metagpt, qian2024chatdev}. 
Despite these successes, coordination remains a critical challenge. As multiple agents interact and contribute to a shared task, ensuring consistent, non-redundant, and well-structured collaboration becomes exceedingly difficult. Recent work has shown that scaling the number of agents does not reliably improve performance, and can in fact degrade results depending on task structure and agent heterogeneity \cite{kim2025towards, pappu2026multi, yang2026understanding, rizvi2025benefits}. In particular, performance deteriorates in settings requiring sequential reasoning or consistent shared states like software repositories \cite{mieczkowski2026language}. Unstructured interaction can further Lead to failures such as hallucinated responsibilities, misinformation, and adversarial behavior among agents \cite{shapira2026agents}.

To overcome these challenges, various frameworks aim to improve team performance with explicitly structured task assignment and interaction.
One common approach is to impose role-based and hierarchical decompositions. Systems like MetaGPT assign fixed functional roles (e.g., Product Manager, Engineer) to agents performing collaborative software tasks \cite{hong2023metagpt}, while ChatDev similarly adopts a fixed hierarchy with role-conditioned specialization  \cite{qian2024chatdev}. Frameworks such as HuggingGPT~\citep{shen2023hugginggpt} similarly use an LLM controller to plan and dispatch heterogeneous expert models against a fixed task taxonomy, but do not adapt the underlying decomposition during execution. More recent approaches introduce feedback-driven re-planning, where a meta-agent iteratively updates plans \cite{liagent}. 

In single-agent settings, reasoning and planning can be structured as chains, graphs, or trees over intermediate steps, or as decompositions into simpler subproblems whose solutions feed forward, improving compositionality and generalization beyond the prompted exemplars \citep{wei2022chain, yao2023tree, besta2024graph, Zhou2022_ue}.

\subsection{Task graphs and coordination protocols}

Recent work identifies parallels between LLM teams and distributed computing systems. Agents with limited local information contend for shared resources, fail without warning, retry, race, and must produce a coherent shared output; notably, this is the exact regime distributed-computing engineers have spent decades modeling \citep{dean2008mapreduce, malewicz2010pregel, moritz2017ray}. 
In both settings, critical tradeoffs arise between different coordination structures \cite{van2017distributed}. 
First, fully centralized architectures can avoid consistency conflicts by designating one Lead to assign tasks and serialize updates, but create bottlenecks and single-points-of-failure. Fully decentralized approaches are more scalable and robust, but agents operating independently on local views of task state can produce conflicting or redundant outputs. 
Second, static approaches commit to a fixed task assignment up front, which simplifies global scheduling but requires complete task visibility in advance. Thus, static assignments cannot adapt when tasks fail, new dependencies emerge, or workloads shift. Alternatively, dynamic approaches assign tasks online as they become available but require mechanisms to maintain consistency as the task evolves.

Task graphs are used in distributed systems (as well as in reinforcement learning and planning~\citep{dietterich2000hierarchical,gopalan2017planning}) to formalize this coordination problem. Nodes represent tasks, edges encode precedence dependencies, and the objective is to schedule tasks to processors efficiently \cite{woo1997task, topcuoglu2002performance}. Classic scheduling algorithms such as HEFT operate at the centralized, static end of both axes, computing a globally optimized assignment before execution begins \cite{topcuoglu2002performance}. Dynamic variants move along the second axis by making assignments online as tasks and dependencies become known \cite{johnson1993concurrent}. Approaches like NABBIT enable both some decentralization and dynamic allocation using a shared task pool and work-stealing, so processors can autonomously claim ready tasks without a central Leader \cite{agrawal2010executing}. Learning-based schedulers such as Decima further adapt policies to workload structure at runtime \cite{mao2019learning}, combining dynamic assignment with learned coordination. Ray similarly operationalizes online task-graph scheduling at scale through a sharded control store and bottom-up distributed scheduler, supporting dynamic task graphs whose structure is not known in advance \citep{moritz2017ray}. Yet Ray's dynamism is in execution scheduling, not task specification: nodes are well-typed remote functions and actor methods registered in advance, and the graph evolves only as those primitives are invoked at runtime. These approaches assume well-defined tasks and explicit control mechanisms, rather than agents that can flexibly and autonomously discover and modify tasks in natural language. These same assumptions persist even when porting over the concept of task graphs to hierarchical decision-making agents~\citep{dietterich2000hierarchical,gopalan2017planning}.

\section{Probabilistic Motivation}
\label{app:prob_motiv}

To motivate the division of labor between Lead and Workers, we develop a probabilistic account of team coordination as posterior inference over task graphs. Since the Lead and Worker agents have varying degrees of visibility, the costs associated with individual steps or rounds of this posterior inference varies between agents. Whereas (approximate) Bayesian inference is commensurate with \emph{thinking}, thinking while balancing the costs of inference induces a higher-level problem of \emph{thinking about how to think} or meta-reasoning~\citep{russell1991principles,hay2012selecting,griffiths2019doing}. We motivate LATTE as encapsulating a meta-reasoning approach to efficiently deploying team-wide computational resources towards the Bayesian inference problem of identifying the best task decomposition conditioned upon all observed data~\citep{raja2007framework,sleight2014multiagent,langlois2020metareasoning}.

\textbf{The inference problem.} Let $G_t$ denote the team's current task decomposition at round $t$, or a hypothesis about how the global task should be decomposed and assigned. We treat the space of valid DAGs as a hypothesis space, and model the team's goal as posterior inference: to find $G_t$ that maximizes $P(G_t \mid D_t)$, where $D_t$ denotes all evidence accumulated in round $t$ (e.g., execution logs, completed outputs, messages, and environment feedback). By Bayes' theorem we have that

\begin{equation}
 \begin{aligned}
P(G_t \mid D_t) = \frac{P(D_t \mid G_t) P(G_t)}{P(D_t)}.
\end{aligned}   
\end{equation}

\noindent The marginal likelihood $P(D_t)=\sum_G P(D_t | G) P(G)$ requires summing over the space of all valid task graphs, which is combinatorially intractable. Additionally, the likelihood $P(D_t | G)$ has no closed form, as $D_t$ is comprised of outputs in natural language or execution traces whose probabilities cannot be evaluated directly. 

\textbf{MCMC as a tractable alternative.}  Markov chain Monte Carlo offers a standard approach to posterior inference that sidesteps computing $P(D_t)$. Rather than evaluating the posterior directly, we construct a Markov chain over task graphs whose stationary distribution is $P(G_t|D_t)$. At each step, a proposal $G'$ is generated and accepted with probability:

\begin{equation}
\begin{aligned}
A(G_t \to G'_t \mid D_t)
&= \min \left( 1, \frac{P(D_t \mid G'_t)\, P(G'_t)\, Q(G_t \mid G'_t, D_t)}{P(D_t \mid G_t)\, P(G_t)\, Q(G'_t \mid G_t, D_t)}\right)
\end{aligned}
\end{equation}

\noindent where $Q(G'_t \mid G_t, D_t)$ is the proposal distribution. Because $P(D_t)$ appears in both numerator and denominator, it cancels exactly, making evaluation of the acceptance ratio feasible without computing the marginal likelihood. 

\textbf{Leader-Worker decomposition as meta-reasoning.} The remaining challenge is constructing a proposal distribution $Q$ that is both tractable and expressive. Observe that one approach would be to take the Lead $\ell$, who has full visibility over the past interactions of all Worker agents $D_t$, and charge them with identifying a new proposal $G'_t$; without regard for the costs of inference, this approach might seem promising. However, this strategy places the onus upon a single agent (in this work, a single LLM) to process a considerable volume of information stored in the full history $D_t$ before proceeding to reason over an exponentially large hypothesis space, both of which degrade LLM response reliability \cite{liu2024lost}. 

Rather than placing all the burdens of inference upon the Lead's shoulders, an alternative and more cost-effective approach would be to empower the individual Worker agents $w_i$. In particular, one could obtain a new task decomposition $G'_{t,i}$ from each Worker $w_i$. From there, one quick option for obtaining a new task decomposition is via simple merging of all Workers' proposals: $G'_t = \texttt{merge}(G'_{t,1}, G'_{t,2}, \ldots)$. Alternatively, each new proposal could be treated as a point estimate in a particle filter and one could simply be chosen uniformly at random. Notably, this type of approach sits at the opposite extreme of the previous Leader-centric approach, where the costs of inference are reduced down to processing the individual local histories of each Worker $d_t^{(i)} \subset D_t$. Unfortunately, as each one lacks global visibility, no one Worker agent is well poised to understand whether or not their proposals actually enhance global task performance for the entire team. 

While the preceding approaches either maximize or sacrifice inference quality and considerably increased or reduced cost, LATTE can obtain a more-efficient solution to the meta-reasoning problem and better balance the quality-cost trade-offs of inference by exploiting the structure of $G_t$. Specifically, we will assign each Worker $w_i$ to a local subtask $g_t^{(i)} \subset G_t$ and accumulates a local execution trace $d_t^{(i)} \subset D_t$. Worker $w_i$ proposes a local update by sampling:

\begin{equation}
\begin{aligned}
g^{(i)\prime}_t \sim Q(\cdot \mid g^{(i)}_t,\, d^{(i)}_t)
\end{aligned}
\end{equation}

\noindent Confining the proposal to $g^{(i)}_t$ keeps each Worker's task tractable since $w_i$ attends to its local trace $d^{(i)}_t$ rather than the full history $D_t$. The candidate global decomposition for round $t$ is then the union of local updates across all Workers $G'_t = \bigcup_{i} g^{(i)\prime}_t$.



The Lead $\ell$ maintains global visibility of $G_t$ and evaluates the acceptance ratio, either commiting $G_{t+1} = G'_t$ or retaining $G_{t+1}=G_t$. Thus, Workers have sufficient local information to propose structural changes within their own scope, while the Lead has the global view needed to evaluate whether a proposal improves the overall decomposition. Critically, this mirrors the Metropolis-Hastings acceptance step~\citep{hastings1970monte}, where the evaluator need only approximate the ratio of unnormalized likelihoods $P(D_t \mid G'_t)P(G'_t) / P(D_t \mid G_t) P(G_t)$ (e.g. a comparison between two specific graphs) rather than integrating over the full hypothesis space; crucially, however, we do not claim that LATTE agents necessarily compute this ratio explicitly. Moreover, LATTE incurs a marginal increase in cost --- for the Lead to assess the benefits of a proposed local update --- beyond the purely local Worker update approach outlined above while capitalizing on the global visibility of the Lead to maintain high-quality inference. Overall, this framing provides a normative account of why the division of labor in LATTE is well-founded and what behaviors the associated LLMs are approximating. We formalize this division of labor as a set of graph mutation operators in the next section.

\section{Graph Mutation Operators}
\label{app:operators_app}

We define seven operators that mutate the task graph; each is invoked by the Lead $\ell$, a Worker $w$, or both. $F_t$ denotes the set of nodes whose dependencies are fully satisfied at time $t$. 

\begin{description}
  \item[$\textsc{Discover}(v, \text{deps})$ $(\ell, w)$]
    Requires $v \notin V_t$, $\text{deps} \subseteq V_t$, and that adding $v$ preserves acyclicity. \\ Adds $v$ to $V_t$, inserts edges from each dependency to $v$ in $E_t$, and initializes \\ $\lambda_t(v) \gets (\bot,
    \texttt{pending})$.

  \item[$\textsc{Assign}(v, w)$ $(\ell)$]
    Requires $\text{status}(v) = \texttt{pending}$ and $w \in \mathcal{W}$.\\
    Sets $\lambda_t(v) \gets (w, \texttt{assigned})$.

  \item[$\textsc{Claim}(v)$ $(w)$]
    Requires $v \in F_t$ and $\text{agent}(v) \in \{\bot, w\}$.\\
    Sets $\lambda_t(v) \gets (w, \texttt{in\_progress})$.

  \item[$\textsc{Complete}(v)$ $(w)$]
    Requires $\text{status}(v) = \texttt{in\_progress}$ and
    $\text{agent}(v) = w$. \\ Sets $\lambda_t(v) \gets (w, \texttt{done})$.

  \item[$\textsc{Release}(v)$ $(\ell)$]
    Requires $\text{status}(v) \in \{\texttt{assigned}, \texttt{in\_progress}\}$.
    \\ Resets $\lambda_t(v) \gets (\bot, \texttt{pending})$.

  \item[$\textsc{Close}(v)$ $(\ell)$]
    Requires $\text{status}(v) \in \{\texttt{assigned}, \texttt{in\_progress}\}$.
    \\ Sets $\lambda_t(v) \gets (\text{agent}(v), \texttt{done})$ without
    requiring the Worker to signal completion.

  \item[$\textsc{Verify}(v)$ $(\ell)$]
    Requires $\text{status}(v) = \texttt{done}$ and $v_{\text{ver}} \notin V_t$. \\ Adds a verification node $v_{\text{ver}}$ to $V_t$, inserts the edge
    $(v, v_{\text{ver}})$ into $E_t$, and \\ initializes $\lambda_t(v_{\text{ver}})
    \gets (\bot, \texttt{pending})$.
\end{description}

\section{LATTE Implementation}
\label{app:latte_impl}

\subsection{Team composition}

Each run consists of one Lead (\textit{Lead}) and $N$ Worker agents (named \textit{Dev1} through \textit{DevN}; $N=4$ in all experiments). All agents run the same underlying model with the same sampling parameters. 

\textbf{Lead agent.}
The Lead runs an isolated planning phase before execution begins. Given the natural-language task description, it has up to 5 turns to produce the initial task graph (nodes + dependency edges) via <discover\_task> actions. During execution the Lead monitors progress, issues <assign\_task> and <release\_task> directives, and can mutate the graph (add nodes, close stale ones). Its context window is capped at the last 10 messages.

\textbf{Worker agents.}
Workers receive task assignments from the Lead and implement them. They can emit <claim\_task>, <complete\_task>, and <discover\_task> to propose adding new subtasks to the graph. Their context window is capped at the last 20 messages. 

\textbf{Concurrency.}
All agents run concurrently within a round — the orchestrator steps the Lead first, then dispatches Workers in parallel, collecting their responses before advancing to the next round.

\subsection{Lead prompt}

You are a senior software engineer leading a team of developers working collaboratively on coding tasks.

Responsibilities:
\begin{enumerate}
    \item Understand the overall project goals and requirements
    \item Break down work and strategically assign tasks to team members
    \item Monitor progress and coordinate the team
    \item Help unblock teammates when they face issues
    \item Review work for quality and consistency
    \item Synthesize results and ensure successful project completion
\end{enumerate}

Work efficiently and delegate appropriately. Trust your teammates to handle their assignments, but provide guidance when needed. Keep communication clear and actionable.

Parallelism:

Teammates can self-assign from the ready queue — they do not need to wait for you. Your job is to keep the graph correct and handle failures, not to manually dispatch every task.

That said, proactively assign tasks when you know a specific agent is the right fit (e.g. after unblocking a straggler, after a verification completes). When several tasks are ready, assign them all at once, one per available agent.

However, be deliberate about what runs in parallel. Avoid assigning two agents to tasks that write to the same function simultaneously. A good rule: tasks that work on distinct functions or files can run in parallel; tasks that both modify the same shared data should be sequenced.

When possible, build a wide graph, not a deep one. Only use `dependencies' to express real implementation ordering (i.e. ``I can't start B until A's output exists''). Don't chain tasks sequentially just for safety — if two tasks touch different functions or files, they can run in parallel with no `dependencies` between them.

With this structure, the moment task-analyze completes, all three implementation tasks become ready in parallel.

Available Actions:

Do NOT edit files yourself — focus on directing your team and verifying their work.

\begin{enumerate}
    \item Assign a task to a teammate: 
    \begin{verbatim}
        <assign_task id="task-1" to="AgentName" />
    \end{verbatim}
    \item Broadcast a message to all teammates (use this to coordinate work):
    \begin{verbatim}
        <broadcast>Your message here</broadcast>
    \end{verbatim}
    \item Run a Python script and see its output:
    \begin{verbatim}
        <run_script path="script.py" />
    \end{verbatim}
    \item Request status from agents:
    \begin{verbatim}
        <request_status />
    \end{verbatim}
    \item Run tests: 
    \begin{verbatim}
        <run_tests />
    \end{verbatim}
    \item Graph updates. The task graph is a living document. Use <discover\_task> to add new tasks whenever: (a) A teammate reports that tests are still failing after completing their task, (b) you notice a dependency was missed or a prior task produced incorrect output, (c) the project needs a verification or integration pass that wasn't planned upfront. 

    Example: if Dev2 finishes implementing a function but broadcasts that tests are still red, add a fix task immediately rather than waiting:
    \begin{verbatim}
    <discover_task id="fix-index" title="Fix index() API bug"
        dependencies="task-2">
        Run <run_tests /> to see failures, fix search_lib.py, 
        confirm all tests pass.
    </discover_task>
    \end{verbatim}
    The index function was implemented with dict input but the tests pass a list.
    \item If a task is high-stakes — it is upstream of many other tasks, or its output is hard to validate later — you can request a verification pass by a second agent:
    \begin{verbatim}
        <verify_task id="task-X" />
    \end{verbatim}  
    This inserts a lightweight review task into the graph that must complete before downstream tasks proceed. The verifying agent will check correctness and fix any issues.
    \item Straggler mitigation. If a teammate has been assigned a task for several rounds without completing it, they may be stuck. Use this action to release the task back to pending so it can be reassigned:
    \begin{verbatim}
        <release_task id="task-X" />
    \end{verbatim}
    This clears the current owner and resets the task to pending. Then reassign it with \begin{verbatim}
        <assign_task id="task-X" to="DevY" />
    \end{verbatim}
    either to a different agent or the same one with clearer instructions. Broadcast a message explaining what the agent should do differently before reassigning.
    \item If the test suite is passing but tasks are still marked "assigned" or "in\_progress" (e.g. a teammate completed the work but forgot to emit <complete\_task>), you can close them directly:
    \begin{verbatim}
        <close_task id="task-X" />
    \end{verbatim}
    Only use this after confirming with <run\_tests /> that tests pass. This is the right action when: all tests are green, a task's work is clearly done in the codebase, and the owning agent is no longer making progress on it.
\end{enumerate}

\subsection{Worker prompt}

You are a skilled software engineer working as part of a development team.

Responsibilities:

\begin{enumerate}
    \item Work on tasks assigned to you by the Lead.

    \item Write clean, well-documented code.

    \item To read an existing file's contents directly, use:
    \begin{verbatim}
<read_file path="math_utils.py" />
    \end{verbatim}
    This returns the file contents immediately --- no script needed. Always prefer this over
    writing a helper script to print a file.

    To execute a script and see its output, use:
    \begin{verbatim}
<run_script path="script.py" />
    \end{verbatim}
    This runs the file and returns stdout/stderr to you. Use this to verify your code works
    before marking a task complete.

    \begin{quote}
    \textbf{Important:} \verb|<run_script>| takes a \texttt{.py} filename only --- it is
    \textbf{not} a shell. Do not pass shell commands like \texttt{ls}, \texttt{head}, or
    \texttt{python3 script.py}. To list files, write a short Python script first with
    \verb|<edit_file>|, then run it with \verb|<run_script>|. Example:
    \end{quote}
    \begin{verbatim}
<edit_file path="check_files.py">
import os; print(os.listdir('.'))
</edit_file>
<run_script path="check_files.py" />
    \end{verbatim}

    \item Use \verb|<run_tests />| to run the test suite and check your work.

    \begin{quote}
    \textbf{Important:} Do not mark a task complete if \verb|<run_tests />| is still failing.
    If you finish your implementation and tests are still red, use \verb|<discover_task>| to
    add a follow-up fix task rather than marking done and moving on. This keeps the problem
    visible to the whole team.
    \end{quote}

    \item Communicate with the team Lead when blocked or in need of clarification.

    \item Complete tasks thoroughly before moving to the next one.
\end{enumerate}

Be proactive, collaborative, and detail-oriented. Focus on producing high-quality work.

\textbf{Discovering New Tasks}

Use \verb|<discover_task>| whenever you uncover work that isn't already in the task list.
When possible, build a wide graph, not a deep one. Only use \texttt{dependencies} to express
real implementation ordering (i.e., ``I cannot start B until A's output exists'').

\begin{verbatim}
<discover_task id="new-task-id" title="Short title" 
dependencies="only-if-truly-required">
  Clear description of what needs to be done and why.
</discover_task>
\end{verbatim}

\subsection{Parameters}

\textbf{Sampling.}
All agents use a temperature of $0.7$ and a maximum output length of $4{,}096$ tokens per call. 

\textbf{Round limits.}
LATTE teams (and baseline teams) were given $40$ rounds total to complete each task. They could complete tasks more quickly by marking tasks as complete (either through the Lead or Workers in a decentralized team). Success was then evaluated based on if their implementations passed the given task's test suite. 

\textbf{Heartbeat monitoring.} We set $H=4$ for all experiments with LATTE. If a Worker was stuck on their subtask implementation for more than 4 rounds without emitting any action, the Lead was notified and prompted to intervene.

\textbf{Claim tie-breaking.} Worker agents were allowed to self-claim tasks from $F_t$ if idle. Concurrent claims were addressed via FIFO by processing order; when multiple Workers claim the same task in the same round, the orchestrator processes agents sequentially and the first claim processed wins, while subsequent claimants get an error message and must re-poll. 

\textbf{API retry.} Anthropic and OpenAI requests rely on their respective SDK retry logic. No per-call wall-clock timeout was set.

\subsection{LATTE Execution Protocol}
\label{app:latte_exec}

\begin{algorithm}[H]
\caption{LATTE Execution}
\label{alg:latte_exec}
\begin{algorithmic}
\Require Task description $\tau$, agents $\mathcal{A} = \{\ell\} \cup \mathcal{W}$, max rounds $T$, heartbeat threshold $H$
\Ensure Final task graph $G_T$

\Statex \textbf{Phase 0: Planning}
\State $G_0 \leftarrow \ell.\textsc{Discover}(\tau)$ \Comment{Leader initializes coordination graph}

\Statex \textbf{Phase 1: Execution}
\For{$t = 1$ \textbf{to} $T$}
    \Statex \quad \textit{// 1. Heartbeat monitoring}
    \State Flag to $\ell$ any $w \in \mathcal{W}$ with no actions in $H$ consecutive rounds

    \Statex \quad \textit{// 2. Frontier identification}
    \State $F_t \leftarrow \{ v \in G_t : \textsc{Status}(v) = \texttt{pending},\ \forall u \in \textsc{Deps}(v):\ \textsc{Status}(u) = \texttt{done} \}$

    \Statex \quad \textit{// 3. Agent dispatching}
    \State Re-engage busy Workers that have received new context since last round
    \State Assign idle Workers to tasks in $F_t$, at most one Worker per task
    \State Invoke $\ell$ if $G_t$ has changed, a heartbeat was flagged, or $\ell$ has been idle for $H$ rounds

    \Statex \quad \textit{// 4. Parallel execution}
    \State All selected agents act in parallel:
    \State \quad $\ell$ receives full graph $G_t$; Workers receive assigned task or $F_t$
    \State \quad Each agent emits actions $\subseteq$ \{\textsc{Discover}, \textsc{Claim}, \textsc{Complete}\}
    \State $G_t \leftarrow \textsc{Apply}(G_{t-1},\ \text{all emitted actions})$

    \Statex \quad \textit{// 5. Termination check}
    \If{$\forall v \in G_t:\ \textsc{Status}(v) = \texttt{done}$}
        \State \Return $G_t$
    \EndIf
\EndFor
\State \Return $G_T$
\end{algorithmic}
\end{algorithm}

\section{Baseline implementations}
\label{app:baselines}
\subsection{MetaGPT}

We use the original paper-release codebase of MetaGPT~\citep{hong2023metagpt}, corresponding to the version publicly available at the time of the ICLR 2024 submission (commit tag \texttt{v0.1}, authored April--August 2023) because it faithfully instantiates the fixed Standard Operating Procedure (SOP) described in the paper.

\textbf{Pipeline.}
MetaGPT structures collaboration as a fixed, sequential SOP over five role-conditioned agents:
\begin{enumerate}
    \item \textbf{ProductManager} (Alice) translates the task description into a Product Requirements Document (PRD), user stories, and a competitive analysis.
    \item \textbf{Architect} (Bob) receives the PRD and produces a system design document, including the Python package name, file structure, and API specifications.
    \item \textbf{ProjectManager} (Eve) reads the system design and issues a task list to the Engineer.
    \item \textbf{Engineer} (Alex, $\texttt{n\_borg}=1$) implements the assigned files sequentially, one file per action, emitting code blocks to shared memory.
    \item \textbf{QaEngineer} (Edward, $\texttt{test\_round\_allowed}=5$) watches for Engineer output and iterates a write-test $\to$ run-code $\to$ debug-error loop up to the allowed round count.
\end{enumerate}

Agents communicate exclusively through a shared publish-subscribe message bus: each role watches a fixed set of upstream action types and acts only when a matching message arrives. The task decomposition, which files to write, what each contains, and in what order, is determined entirely during the planning phases (steps 1--3) and cannot be revised during execution.

\textbf{Parameters.}
All agents use the same model and sampling parameters as the LATTE and baseline runs described in Appendix~\ref{app:latte_impl}. Each MetaGPT run is allocated \texttt{n\_round}$=40$ total team rounds to match the other team conditions. The QaEngineer is initialized with \texttt{test\_round\_allowed}$=5$, matching the upper bound of debug-and-fix cycles in the original paper.



\emph{Mismatch between SOP outputs and task evaluation.}
Because each role in the sequential pipeline operates on the prior role's output rather than the shared task environment, artifacts accumulate in locations determined by the Architect's upfront design rather than by where the task expects them. 
More broadly, because the task decomposition is committed before any code is executed, the Engineer cannot discover latent task structure such as which functions share internal helpers, which bugs are actually present in a given file, or which data columns carry the signal of interest. On nonstationary tasks where the correct decomposition only becomes apparent during execution, the upfront plan is systematically misspecified, and the SOP provides no mechanism to revise it.

\subsection{Leader-Worker Hierarchies}

Leader-Worker teams were implemented using a lightweight orchestrator with no task-graph infrastructure. Agents could edit files, run scripts, run tests, and broadcast messages to teammates. Unlike the graph-based conditions, there was no planning phase and no task state. Coordination relied entirely on the Lead broadcasting directions and
teammates editing files in response.

At the start of each round the orchestrator appended the latest test output to every agent's context window. The Lead then ran first, followed by all $N$ teammates in parallel. After each round the orchestrator ran the test suite internally to detect success. A run was marked successful when all tests passed and the Lead certified task completion. 

\paragraph{Lead prompt}

You are a senior software engineer leading a team of developers working collaboratively on coding tasks.

Responsibilities:
\begin{enumerate}
    \item Understand the overall project goals and requirements
    \item Break down work and strategically assign tasks to team members
    \item Monitor progress and coordinate the team
    \item Help unblock teammates when they face issues
    \item Review work for quality and consistency
    \item Synthesize results and ensure successful project completion
\end{enumerate}

Work efficiently and delegate appropriately. Trust your teammates to handle their assignments, but provide guidance when needed. Keep communication clear and actionable.

Available Actions:

Do NOT edit files yourself — focus on directing your team and verifying their work.

\begin{enumerate}
    \item Broadcast a message to all teammates (use this to coordinate work):
    \begin{verbatim}
        <broadcast>Your message here</broadcast>
    \end{verbatim}
    \item Run a Python script and see its output:
    \begin{verbatim}
        <run_script path="script.py" />
    \end{verbatim}
\end{enumerate}

\paragraph{Worker prompt}

You are a skilled software engineer working as part of a development team.

Responsibilities:

\begin{enumerate}
    \item Work on tasks assigned to you by the Lead.

    \item Write clean, well-documented code.

    \item To read an existing file's contents directly, use:
    \begin{verbatim}
<read_file path="math_utils.py" />
    \end{verbatim}
    This returns the file contents immediately --- no script needed. Always prefer this over
    writing a helper script to print a file.

    To execute a script and see its output, use:
    \begin{verbatim}
<run_script path="script.py" />
    \end{verbatim}
    This runs the file and returns stdout/stderr to you. Use this to verify your code works
    before marking a task complete.

    \begin{quote}
    \textbf{Important:} \verb|<run_script>| takes a \texttt{.py} filename only --- it is
    \textbf{not} a shell. Do not pass shell commands like \texttt{ls}, \texttt{head}, or
    \texttt{python3 script.py}. To list files, write a short Python script first with
    \verb|<edit_file>|, then run it with \verb|<run_script>|. Example:
    \end{quote}
    \begin{verbatim}
<edit_file path="check_files.py">
import os; print(os.listdir('.'))
</edit_file>
<run_script path="check_files.py" />
    \end{verbatim}

    \item Use \verb|<run_tests />| to run the test suite and check your work.

    \begin{quote}
    \textbf{Important:} Do not mark a task complete if \verb|<run_tests />| is still failing.
    \end{quote}

    \item Communicate with the team lead when blocked or in need of clarification.

    \item Complete tasks thoroughly before moving to the next one.
\end{enumerate}

Be proactive, collaborative, and detail-oriented. Focus on producing high-quality work.

\subsection{Decentralized Teams}

Decentralized teams used the same lightweight orchestrator and action vocabulary as the Leader-Worker condition, but with no designated Leader. To hold agent count constant across conditions, we instantiated $N{+}1$ symmetric peer agents (matching the $1$ Lead $+ N$ Worker headcount of the other conditions). All agents ran in parallel every round with no sequential ordering; coordination relied solely on broadcast messages. As in the Leader-Worker condition, the latest test output was appended to every agent's context at the start of each round, and success was detected by running the test suite internally and any agent marking task completion.

\paragraph{Peer prompt}

You are a skilled software engineer working as part of a development team.

Responsibilities:

\begin{enumerate}
    \item Work on your tasks efficiently and effectively

    \item Write clean, well-documented code

    \item To read an existing file's contents directly, use:
    \begin{verbatim}
<read_file path="math_utils.py" />
    \end{verbatim}
    This returns the file contents immediately --- no script needed. Always prefer this over
    writing a helper script to print a file.

    To execute a script and see its output, use:
    \begin{verbatim}
<run_script path="script.py" />
    \end{verbatim}
    This runs the file and returns stdout/stderr to you. Use this to verify your code works
    before marking a task complete.

    \begin{quote}
    \textbf{Important:} \verb|<run_script>| takes a \texttt{.py} filename only --- it is
    \textbf{not} a shell. Do not pass shell commands like \texttt{ls}, \texttt{head}, or
    \texttt{python3 script.py}. To list files, write a short Python script first with
    \verb|<edit_file>|, then run it with \verb|<run_script>|. Example:
    \end{quote}
    \begin{verbatim}
<edit_file path="check_files.py">
import os; print(os.listdir('.'))
</edit_file>
<run_script path="check_files.py" />
    \end{verbatim}

    \item Use \verb|<run_tests />| to run the test suite and check your work.

    \item Communicate with your teammates as needed.

    \item Complete tasks thoroughly before moving to the next one.
\end{enumerate}

Be proactive, collaborative, and detail-oriented. Focus on producing high-quality work.

\subsection{Static Graph Ablation}

The static condition used the same graph-based orchestrator as the dynamic condition, but with task discovery and reassignment disabled after the planning phase. During planning, the Lead received a prompt instructing it to emit a complete task graph upfront, specifying every task and its dependencies before execution began. Once planning concluded, the graph was frozen: mid-run task discovery (\texttt{discover\_task}), straggler release (\texttt{release\_task}),
verification insertion (\texttt{verify\_task}), and automatic fix-task injection on test failure were all disabled at the orchestrator level. Teammates could not self-assign; the Lead was responsible for assigning every task at the start of execution via \texttt{assign\_task}. All $N$ teammates were dispatched every
round regardless of task availability. If a teammate failed to complete an assigned task, no recovery mechanism was available.

\subsection{Lead prompt}

You are a senior software engineer leading a team of developers working collaboratively on coding tasks.

\textbf{Responsibilities}
\begin{enumerate}
    \item Understand the overall project goals and requirements
    \item Break down work and strategically assign tasks to team members upfront
    \item Monitor progress and answer questions from teammates
    \item Synthesize results and ensure successful project completion
\end{enumerate}

Work efficiently and delegate appropriately. Trust your teammates to handle their assignments.

\textbf{Parallelism}

Teammates do not self-assign — you must assign every task. Assign all ready tasks at the start, distributing work evenly across available agents.

Be deliberate about what runs in parallel. Avoid assigning two agents to tasks that write to the same function simultaneously. A good rule: tasks that work on distinct functions or files can run in parallel; tasks that both modify the same shared data should be sequenced.

When possible, build a wide graph, not a deep one. Only use `dependencies' to express real implementation ordering (i.e. ``I can't start B until A's output exists''). Don't chain tasks sequentially just for safety — if two tasks touch different functions or files, they can run in parallel with no `dependencies` between them.
With this structure, the moment task-analyze completes, all three implementation tasks become ready in parallel.

\textbf{Available Actions}

Do NOT edit files yourself — focus on directing your team and verifying their work.

\begin{enumerate}
    \item Assign a task to a teammate: 
    \begin{verbatim}
        <assign_task id="task-1" to="AgentName" />
    \end{verbatim}
    \item Broadcast a message to all teammates (use this to coordinate work):
    \begin{verbatim}
        <broadcast>Your message here</broadcast>
    \end{verbatim}
    \item Run a Python script and see its output:
    \begin{verbatim}
        <run_script path="script.py" />
    \end{verbatim}
    \item Request status from agents:
    \begin{verbatim}
        <request_status />
    \end{verbatim}
    \item Run tests: 
    \begin{verbatim}
        <run_tests />
    \end{verbatim}
    \item Fixed plan. The task decomposition and assignments are fixed at the start. Once tasks are assigned, they run to completion without reassignment or modification. Your job is to: (a) Assign all tasks upfront based on the task graph and agent availability. (b) Answer teammates' questions via broadcast if they get stuck. (c) Run tests at the end to confirm completion.
    
    You cannot release tasks, reassign workers, or insert new tasks once execution has begun. If a teammate fails to complete a task, the team absorbs that outcome — do not attempt to recover by reassigning.
\end{enumerate}

\subsection{Worker prompt}

You are a skilled software engineer working as part of a development team.

\textbf{Responsibilities}

\begin{enumerate}
    \item Work on tasks assigned to you by the Lead.

    \item Write clean, well-documented code.

    \item To read an existing file's contents directly, use:
    \begin{verbatim}
<read_file path="math_utils.py" />
    \end{verbatim}
    This returns the file contents immediately --- no script needed. Always prefer this over
    writing a helper script to print a file.

    To execute a script and see its output, use:
    \begin{verbatim}
<run_script path="script.py" />
    \end{verbatim}
    This runs the file and returns stdout/stderr to you. Use this to verify your code works
    before marking a task complete.

    \begin{quote}
    \textbf{Important:} \verb|<run_script>| takes a \texttt{.py} filename only --- it is
    \textbf{not} a shell. Do not pass shell commands like \texttt{ls}, \texttt{head}, or
    \texttt{python3 script.py}. To list files, write a short Python script first with
    \verb|<edit_file>|, then run it with \verb|<run_script>|. Example:
    \end{quote}
    \begin{verbatim}
<edit_file path="check_files.py">
import os; print(os.listdir('.'))
</edit_file>
<run_script path="check_files.py" />
    \end{verbatim}

    \item Use \verb|<run_tests />| to run the test suite and check your work.

    \begin{quote}
    \textbf{Important:} Do not mark a task complete if \verb|<run_tests />| is still failing.
    If you finish your implementation and tests are still red, use \verb|<discover_task>| to
    add a follow-up fix task rather than marking done and moving on. This keeps the problem
    visible to the whole team.
    \end{quote}

    \item Communicate with the team lead when blocked or in need of clarification.

    \item Complete tasks thoroughly before moving to the next one.
\end{enumerate}

Be proactive, collaborative, and detail-oriented. Focus on producing high-quality work.

\section{Experiments}
\label{app:experiments}

\subsection{Setup}

\textbf{Task 1: Exploratory data analysis} Identifying meaningful patterns in a dataset is inherently time-consuming and open-ended. The process begins with preprocessing and filtering, which can be handled by a single agent. Subsequently, multiple agents can explore the data in parallel along diverse directions (e.g., characterizing distributions, identifying outliers, generating visualizations). A final agent can then aggregate and synthesizes these findings. Importantly, purely static task decompositions are likely to be suboptimal because the underlying structure of the data and promising directions only emerge during analysis.

We first simulated a tabular HR dataset containing 400 employee records with eight deliberately opaque column names and no data dictionary. Three ground-truth properties were planted in the data: (1) employee satisfaction score is the dominant churn predictor, (2) salary follows a bimodal distribution reflecting two workforce tiers, and (3) churn rate varies substantially by department. Agents were tasked with producing three interdependent artifacts: a \texttt{config.py} establishing the shared column-name mapping and feature schema (the foundation all downstream scripts must import), a structured \texttt{findings.json} with quantitative supporting evidence across five analytic categories (distributions, relationships, subgroup effects, outliers, and missing data), and a written \texttt{summary.txt} narrative of at least 100 words. Correctness is evaluated by a private test suite that checks whether the true churn column was identified, whether the column--feature classification matches the data types, and whether each of the three planted properties appears in the findings with correct directional claims.

\textbf{Task 2: Debugging} 
Debugging is another task which is inherently nonstationary; agents must run tests, read outputs, change the code to solve potentially multiple errors, and repeat the process until the problems are solved. It also rewards both parallelism and consistency, since multiple agents can simultaneously diagnose independent errors, yet some errors require changes to composite functions that can only be verified after dependencies are fixed. In Task 2, we planted 8 bugs (1 per function) in a Python signal-processing library covering a range of common numerical mistakes, such as a $+1$ in the normalization denominator or inequality errors. Unlike Task 1, in which performance was evaluated after the agents marked task completion, the team had access to the full test suite. Success required all tests in the public test suite to pass. 

\textbf{Task 3: Library extension} 
Finally, code generation is a task that is inherently time consuming; teams of agents can improve performance by working in parallel to generate independent parts of a codebase. 
In this task, agents were given a partially implemented Python text-processing library with three working files: a \texttt{Document} class, \texttt{Tokenizer} class, and \texttt{utils} module. They must extend the existing classes and build six new modules from stub files: \texttt{sentiment.py}, \texttt{keywords.py}, \texttt{summarizer.py}, \texttt{similarity.py}, \texttt{formatter.py}, and \texttt{pipeline.py}. Each stage of generation required different numbers of agents: two existing classes must be extended first, which only required two agents to change; then, the new modules could be generated independently in parallel, and finally only one agent needed to integrate everything into the final pipeline. Agents were encouraged to write their own tests, and correctness was determined after they mark test completion by a private test suite which tests common outputs and edge cases. 

\subsection{Trials}

For each base model, we ran a total of 150 trials (3 tasks, 5 team structures, 10 repetitions). For Claude Sonnet 4-6, this amounted to 2,136 API cells, 40.7M input tokens, and 8.5M output tokens (approximately \$250). For GPT 5.2, this amounted to 4,582 API calls, 83.1M input tokens, 14M output tokens (approximately \$260). 

\subsection{Planning Overhead Analysis}
\label{app:graph_overhead}

LATTE's adaptivity introduces two components of orchestration cost. First, an initial planning phase in which the Lead seeds the task graph: dynamic conditions seed a mean of 3.8 ($\pm 2.0$) nodes in 14.5s, compared to 8.3 ($\pm 1.8$) in 16.7s for static, which front-loads the full decomposition. Second, in-execution orchestration: the Lead remains active in, on average, 30.8\% of of execution rounds under dynamic versus 14.8\% for static, adding a mean of 0.9 extra lead-active rounds per run.

While these are real costs, they do not outweigh the benefits that LATTE provides to teams for dynamic and complex tasks. Across all tasks and models, dynamic conditions complete in a mean of 9.8 rounds and 148K tokens, compared to 15.9 rounds and 297K tokens for static. Thus, each extra lead-active round of dynamic overhead yields approximately six fewer total rounds of Worker execution.

\end{document}

%% file: commands.tex
\definecolor{dblue}{RGB}{98, 140, 190}
\definecolor{dlblue}{RGB}{216, 235, 255}
\definecolor{dgreen}{RGB}{124, 155, 127}
\definecolor{dpink}{RGB}{207, 166, 208}
\definecolor{dyellow}{RGB}{255, 248, 199}
\definecolor{dgray}{RGB}{46, 49, 49}

\newcommand{\durl}[1]{\textcolor{dblue}{\underline{\url{#1}}}}

\newmdenv[
  topline=false,
  bottomline=false,
  rightline = false,
  leftmargin=10pt,
  rightmargin=0pt,
  innertopmargin=0pt,
  innerbottommargin=0pt
]{innerproof}

\newcounter{DaveDefCounter}
